\RequirePackage{amsmath}
\documentclass[epj]{svjour}

\usepackage{amssymb}
\usepackage{amsmath}
\usepackage{braket}
\usepackage{siunitx}

\usepackage{lineno}

\usepackage{cuted}
\usepackage{graphicx}
\usepackage{siunitx}
\usepackage{multirow}
\usepackage{mathtools}

\usepackage{pbox}

\begin{document}
\title{High-precision half-life determination of $^{14}$O via direct $\beta$~counting}
\author{
S.~Sharma\inst{1} \and 
G.F.~Grinyer\inst{1}\thanks{\emph{Corresponding Author: Gwen.Grinyer@uregina.ca} } \and
G.C.~Ball\inst{2} \and
J.R.~Leslie\inst{3} \and
C.E.~Svensson\inst{4} \and
F.A.~Ali\inst{4}\thanks{\emph{Present address: Department of Physics, College of Education, University of Sulaimani, P.O. Box 334, Sulaimani, Kurdistan Region, Iraq}} \and
C.~Andreoiu\inst{5} \and
N.~Bernier\inst{2,6}\thanks{\emph{Present address: Department of Physics and Astronomy, University of the Western Cape, P/B X17, Belleville 7535, South Africa and Department of Physics and Engineering, University of Zululand, P/B X1001, KwaDlangezwa 3886, South Africa}} \and
S.S.~Bhattacharjee\inst{2} \and
V.~Bildstein\inst{4} \and
C.~Burbadge\inst{4} \and
R.~Caballero-Folch\inst{2} \and
R.~Coleman\inst{4} \and
A.~Diaz~Varela\inst{4} \and
M.R.~Dunlop\inst{4} \and
R.~Dunlop\inst{4} \and
A.B.~Garnsworthy\inst{2} \and
E.~Gyabeng~Fuakye\inst{1} \and
G.M.~Huber\inst{1} \and
B.~Jigmeddorj\inst{4} \and
K.~Kapoor\inst{1} \and
A.T.~Laffoley\inst{4} \and
K.G.~Leach\inst{7} \and
J.~Long\inst{8} \and
A.D.~MacLean\inst{4} \and
C.R.~Natzke\inst{2,7} \and
B.~Olaizola\inst{2}\thanks{\emph{Present address: ISOLDE, CERN, CH-1211 Geneva 23, Switzerland}} \and
A.J.~Radich\inst{4} \and
N.~Saei\inst{1} \and
J.T.~Smallcombe\inst{2}\thanks{\emph{Present address: Oliver Lodge Laboratory, University of Liverpool, Liverpool, L69 7ZE, UK}} \and
A.~Talebitaher\inst{1} \and
K.~Whitmore\inst{5} \and
T.~Zidar\inst{4}
}
\institute{
Department of Physics, University of Regina, Regina, SK S4S 0A2, Canada \and
TRIUMF, 4004 Wesbrook Mall, Vancouver, BC V6T 2A3, Canada \and
Department of Physics, Queen's University, Kingston, ON K7L 3N6, Canada \and
Department of Physics, University of Guelph, Guelph, ON N1G 2W1, Canada \and
Department of Chemistry, Simon Fraser University, Burnaby, BC V5A 1S6, Canada \and
Department of Physics and Astronomy, University of British Columbia, Vancouver, BC  V6T 1Z4, Canada \and
Department of Physics, Colorado School of Mines, Golden, CO 80401, USA \and
Department of Physics, University of Notre Dame, Notre Dame, IN 46556, USA
}

\date{Received: date / Revised version: \today}
\abstract{
The half-life of the superallowed Fermi $\beta^+$~emitter $^{14}$O was determined to high precision via a direct $\beta$~counting experiment performed at the Isotope Separator and Accelerator (ISAC) facility at TRIUMF.  The result, $T_{1/2}$($^{14}$O)~=~70619.2(76)~ms, is consistent with, but is more precise than, the world average obtained from 11 previous measurements.  Combining the $^{14}$O half-life deduced in the present work with the previous most precise measurements of this quantity leads to a reduction in the overall uncertainty, by nearly a factor of 2.  The new world average is $T_{1/2}$($^{14}$O)~=~70619.6(63)~ms with a reduced $\chi^2$~value of 0.87 obtained from 8 degrees of freedom. 
}

\maketitle

\section{Introduction}
High precision measurements of the $ft$~values for superallowed Fermi $\beta$~decays are crucial for providing fundamental tests of the Standard Model description of electroweak interactions~\cite{Har20}. All nuclear $\beta$~decays are characterized by their $ft$~value, which depends on three quantities that can each be determined experimentally: the energy difference between the parent and daughter states involved in the $\beta$~transition~($Q$~value), the branching ratio~($BR$), or fraction of parent decays that populate the particular daughter state of interest, and the half-life~($T_{1/2}$) of the parent nucleus.

For the specific case of superallowed Fermi $\beta$~decay between isospin $T$~=~1 isobaric analogue states, the $ft$~value, once corrected for radiative and isospin-symmetry breaking effects, can be written as a corrected $\mathcal{F}t$~value according to the expression~\cite{Har20}:
\begin{eqnarray}
\mathcal{F}t = \textit{ft}(1+\delta_R^\prime)(1 + \delta_{NS} -\delta_C)= \frac{K}{2G_V^2(1+\Delta^V_R)},
\label{Ft_corrected}
\end{eqnarray}
where $\delta_R^\prime$, $\delta_{NS}$, $\delta_{C}$ and $\Delta_R^V$ are all theoretically calculable correction factors that account for radiative and transition dependent electromagnetic effects in addition to isospin-symmetry breaking, $\delta_C$, that depends sensitively on the structure of the parent and daughter states. These corrections are small in magnitude ($\leq$~2.5$\%$) and are described in detail in Ref.~\cite{Har20}.  The term $G_V$ in Eqn.~\ref{Ft_corrected} is the vector coupling constant for semi-leptonic weak interactions and $K$~is a group of well-known and fundamental physical constants with a numerical value that evaluates to $K/{(\hbar c)^6}$~=~8120.27648(26)$\times$10$^{-10}$~GeV$^{-4}$s~\cite{Har20}.  From Eqn.~\ref{Ft_corrected}, every experimentally determined $ft$~value can be used to independently determine a value for $G_V$.  According to the conserved vector current (CVC) hypothesis~\cite{Fey58}, this value should be the same for all transitions and implies that the $\mathcal{F}t$ values themselves should also be constant, within their quoted uncertainties.      

The constancy of the $\mathcal{F}t$~values, and thus the validity of the CVC hypothesis itself, has been firmly established at the level of 9$\times$10$^{-5}$ using a set of 15 precisely determined $T$~=~1 superallowed Fermi $\beta$-decay transitions~\cite{Har20}.  These decays, particularly the lightest superallowed transitions, are also used to set strict limits on the possible existence of a scalar term in the weak interaction~\cite{Har20}.  A scalar interaction, whether fundamental or induced by the vector part, would result in a deviation to the constancy of the $\mathcal{F}t$~values that would be inversely proportional to the average decay energy $\langle$1/$W\rangle$.  Since this quantity is largest for decays of $^{10}$C and $^{14}$O, the $ft$~values for these decays are essential for setting limits on scalar currents~\cite{Dun16}.

In this work, we focus on the superallowed decay of $^{14}$O, which is already one of the most precisely determined $ft$~values.  Its half-life,  $T_{1/2}$~=~70619(11)~ms~\cite{Har20}, is currently obtained from a weighted average of 11 previous measurements.  The $Q$-value for this decay was recently measured to high-precision in Ref.~\cite{Val15} and the resulting world average, $Q$~=~2831.543(76)~keV~\cite{Har20}, is nearly an order of magnitude more precise than the value adopted in the previous survey of the superallowed data~\cite{Har15}.  The recommended value for the superallowed branching ratio to the 0$^+$ isobaric analogue state at 2.31~MeV excitation in the daughter nucleus, $^{14}$N, is $BR$~=~99.446(13)$\%$~\cite{Har20}. Combining the above quantities with $P_{EC}$~=~0.088$\%$, that accounts for electron capture, the experimental $ft$~value for the superallowed decay of $^{14}$O is  $ft$~=~3042.23(84)~s~\cite{Har20}, a result that has been determined at the level of $\pm$~0.027$\%$ precision.  

A new, and even more precise, value for the half-life of $^{14}$O is presented that was deduced from an experiment performed at the Isotope Separator and Accelerator (ISAC) facility at TRIUMF.  This new result represents the most precise half-life ever reported for this decay and is, by itself, more precise than the world average obtained from all 11 previous measurements.

\section{Experiment}

The experiment was performed at TRIUMF, Canada's particle accelerator centre.  Radioactive ion beams are produced at the ISAC facility using the Isotope Separation On-Line (ISOL) technique.  The 480~MeV primary proton beam from the TRIUMF main cyclotron was impinged at 50~$\mu$A intensity onto a thick silicon-carbide (11.97~g/cm$^2$ Si) production target.  Elements produced from spallation reactions, diffused and effused from the target and entered a coupled Forced Electron Beam Ion Arc Discharge (FEBIAD) ion source~\cite{Ame14} where they were ionized.  Ions were extracted from the source at 40~keV and were sent through a low-resolution magnetic pre~separator and a high-resolution ${\Delta}m/m$~=~1/2500 mass separator to select the isotopes of interest.  The production yield of singly ionized $^{14}$O$^+$ was measured to be an order of magnitude less than carbon-monoxide $^{12}$C$^{14}$O$^+$, presumably due to the presence of carbon in the SiC production target that facilitated the formation of CO~molecules.  Since $^{12}$C is stable, its presence did not have any impact on the $^{14}$O decay counting measurement.  The mass separator was thus tuned to mass $A$~=~26 to select the $^{12}$C$^{14}$O$^+$ ions and take advantage of the higher $^{14}$O yield at this setting.  An average intensity of 10$^5$~molecules/s was delivered to the experiment over a period of approximately 4~days.  

The $^{12}$C$^{14}$O$^+$ beam was implanted into a 17.2~$\mu$m thick aluminum layer of an aluminized mylar tape.  After implanting the beam for a set period of time to collect a sufficient amount of activity, the beam was turned off.  This was followed by a pause of 22.2~s in order to give sufficient time for any possible short-lived beam contaminants to decay before transporting the remaining $^{14}$O sample to the $\beta$~counter.  The detector used in this experiment was a gas-filled proportional counter that has been described in detail and used in several previous high-precision half-life measurements~\cite{Dun16,Bal01,Gri05,Gri08,Fin11}. 

Signals from the gas counter were amplified using an Ortec~579 amplifier whose output signals were then sent to an Ortec~436 constant fraction discriminator~(CFD).  The CFD was used to set and adjust an energy threshold to trigger on the $\beta$~particles emitted from the radioactive decays of nuclei in the sample.  Logic pulses from the output of the CFD were then duplicated in a fan-in-fan-out module and were sent to two inputs of a LeCroy~222 dual gate and delay generator.  This module was used to generate two distinct fixed and non-extendible dead times that were applied to every trigger event.  These were determined to be 2.9958(42)~$\mu$s and 4.0105(43)~$\mu$s using the source-plus-pulser technique~\cite{Bae65}.  Two independent multi-channel scalers (MCS), referred to as the old and new MCS, were then used to bin the decay data from the counter into two separate data streams.

\subsection{Detector Setup}

\begin{figure}[tp]
\centering
\rotatebox{0}{\includegraphics[width=220pt]{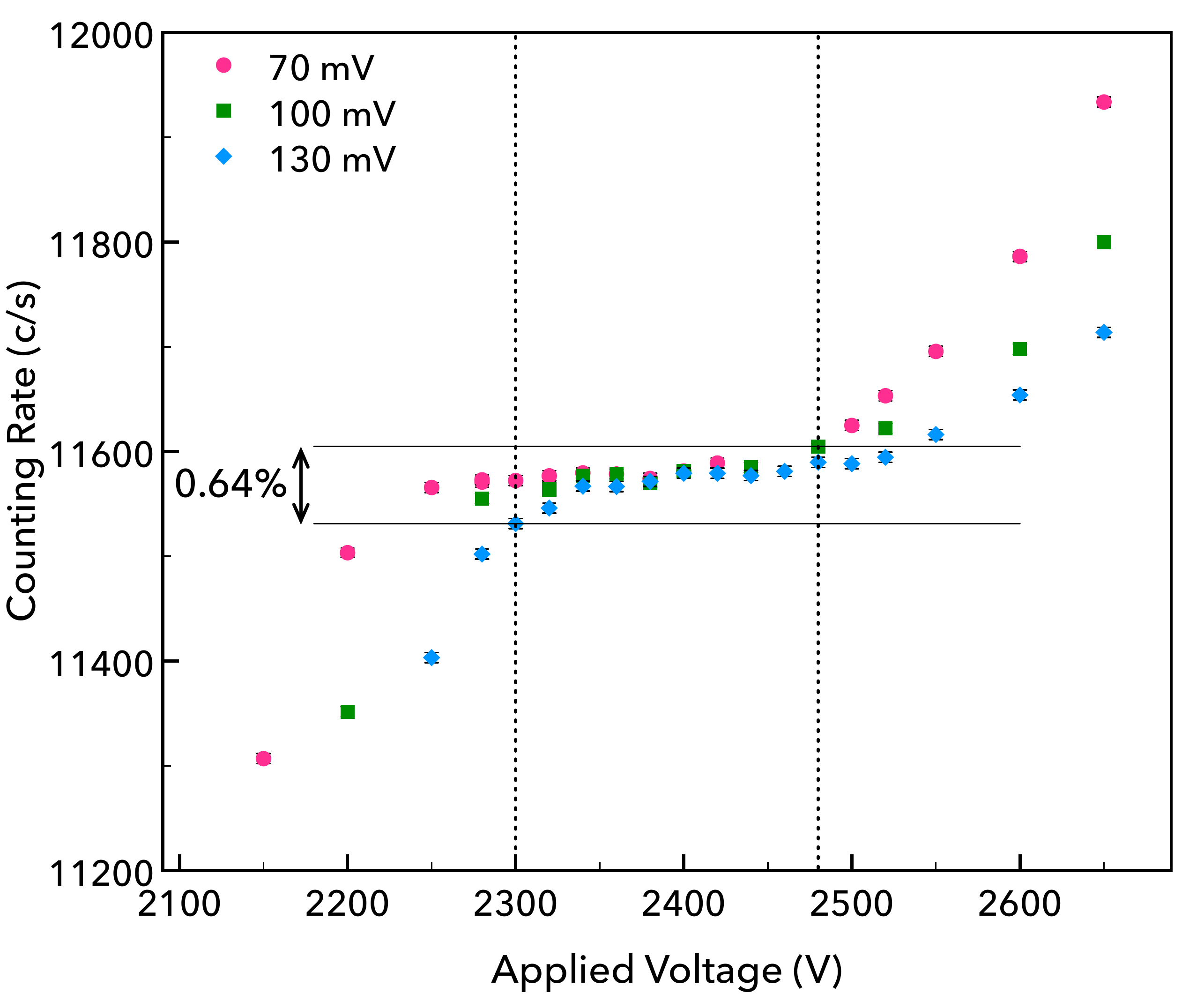}}
\caption{(Color online) Measured count rate of a $^{90}$Sr calibration source as a function of applied voltage for different discriminator threshold settings.  The plateau region was determined to be between 2300~V to 2480~V, as indicated by the vertical bands.  The maximum variation in counting rate observed in this region was less than 0.64$\%$ (horizontal bands). }
\label{Plateau}
\end{figure}

The gas counter was operated in the voltage plateau of the proportional region to ensure that the counting rate is nearly independent of the gain.  The plateau region was determined by recording the count rate of a long lived $^{90}$Sr source as a function of the applied bias voltage.  As the plateau region should also not depend on the discriminator setting, several measurements were performed at CFD threshold values of 70~mV, 100~mV and 130~mV.  From the data presented in Fig.~\ref{Plateau}, the proportional region of the detector was determined to be between 2300~V and 2480~V (inclusive) and, as shown, is largely independent of the discriminator thresholds.

\subsection{Data Selection Criteria}

Decay data for the $^{14}$O experiment were collected in cycles that consisted of a beam on (collection) period, a tape move to position the $^{14}$O activity at the center of the gas counter, a ``cool down" period to allow for any short-lived contaminants in the beam to decay away, and then the counting measurement to record the exponential decay activity.  The precise values of the collection time, tape move and decay measurement were varied on a cycle-by-cycle basis.  In general, one complete cycle typically lasted around 27~minutes, with the total of the collection, ``cool down" and tape movement times taking at most 3~minutes to complete.  The long decay times of approximately 24~minutes for each cycle were necessary to observe the entire decay of the $^{14}$O samples and to determine the resulting background level afterwards, that is treated as a free parameter and potentially contains important information about long-lived contaminants.  Over the course of the 4-day experiment, a total of 142 decay cycles were recorded.

The first selection criterion that was applied to the experimental data was to reject those cycles that had very few, or even zero, total counts recorded during the decay measurement.  These corresponded to cycles whose activity collection period overlapped with an interruption in the ISAC beam delivery.  A total of 17 cycles were removed from the analysis based on this selection criterion.  

For the remaining cycles, a second selection was performed by analyzing the reduced $\chi^2$~values determined from fitting the experimental dead-time corrected decay curves to a single exponential decay plus a free constant background.  With the exception of one cycle, the reduced $\chi^2$ values ranged between 0.81 and 1.27.  The one cycle that was identified in this analysis, due to its anomalously large $\chi^2$~value, had a significant excess of counts in the middle of the decay counting period.  This excess lasted only seconds in duration and was likely due to a burst of electronics noise that was generated close to the detector or picked up in the electronics chain.  This cycle was therefore removed from the analysis.

A total of 124 cycles, of the 142 cycles collected, passed these two selection criteria and were used in the final analysis to determine the half-life of $^{14}$O.

\subsection{Analysis} 

\begin{figure}[tp]
\centering
\rotatebox{0}{\includegraphics[width=220pt]{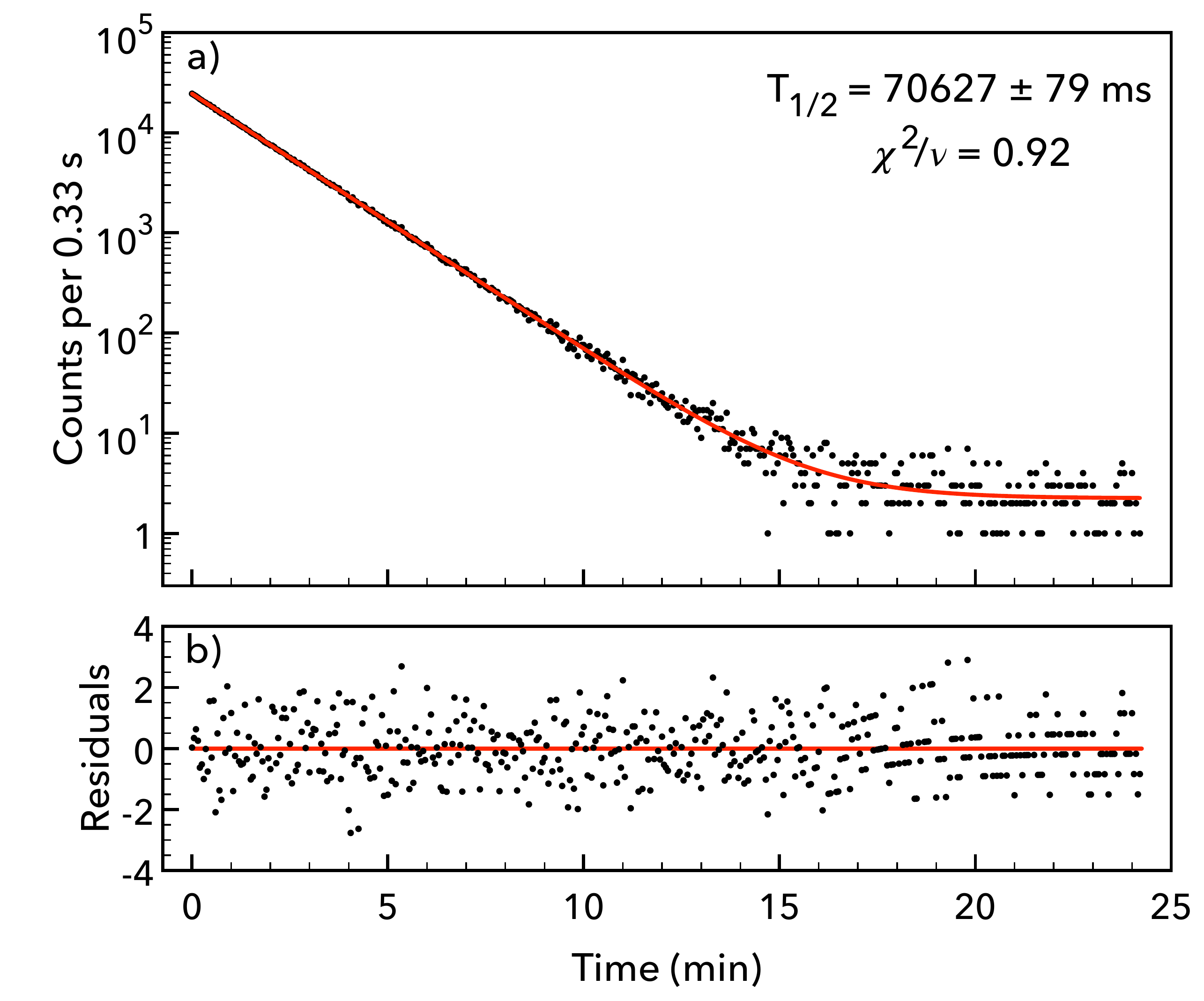}}
\caption{(Color online) a) Dead-time corrected $^{14}$O decay data (points) and resulting fit (line) for one cycle: 140~s collection, 24.3~min decay, 70~mV threshold and 2450~V detector bias. b) Residuals (points, ($y_i$-$y_{fit}$)/$\sigma_i$) obtained from the fit to the experimental data in a) and zero (line) to guide the eye.  For clarity, statistical uncertainties on the data are not shown.}
\label{Run309}
\end{figure}

The 124 cycles that passed the initial selection criteria were corrected for dead-time losses using the measured fixed and non-extendible dead times according to the procedure described in Ref.~\cite{Kos97} and fit using a maximum-likelihood technique~\cite{Gri05}.  The fit function considered only the exponential decay of $^{14}$O and a constant background according to the expression: 
\begin{eqnarray}
y_{fit}(t) = a_1e^{-a_2t} + a_3,
\label{fit_function}
\end{eqnarray}
where $a_1$ is the initial activity of the $^{14}$O sample, $a_2$ is the $^{14}$O decay constant, and $a_3$ the constant background.  All were treated as free parameters in the fit.  A sample decay curve and resulting fit for a single decay cycle is presented in Fig.~\ref{Run309}.  Additional parameters were added to the function to search for, and set limits on, the possibility of trace amounts of beam contaminants.  This analysis is described below in Sec.~\ref{isobaric_systematic}. 

\section{Results}

The half-life of $^{14}$O was determined from the fits to the individual cycles according to Eqn.~\ref{fit_function}.  For every cycle, the half-life was deduced from the decay data recorded in each of the MCS modules.  The average values obtained from all 124 cycles were $T_{1/2}$~=~70619.0(70)~ms from the new MCS and $T_{1/2}$~=~70619.5(70)~ms from the old MCS.  These results are not statistically independent because they were obtained from the same input data recorded by two separate modules.  Because they are in excellent agreement with each other, a simple (non-weighted) average was adopted as the final result.

\begin{figure}[tp]
\centering
\rotatebox{0}{\includegraphics[width=220pt]{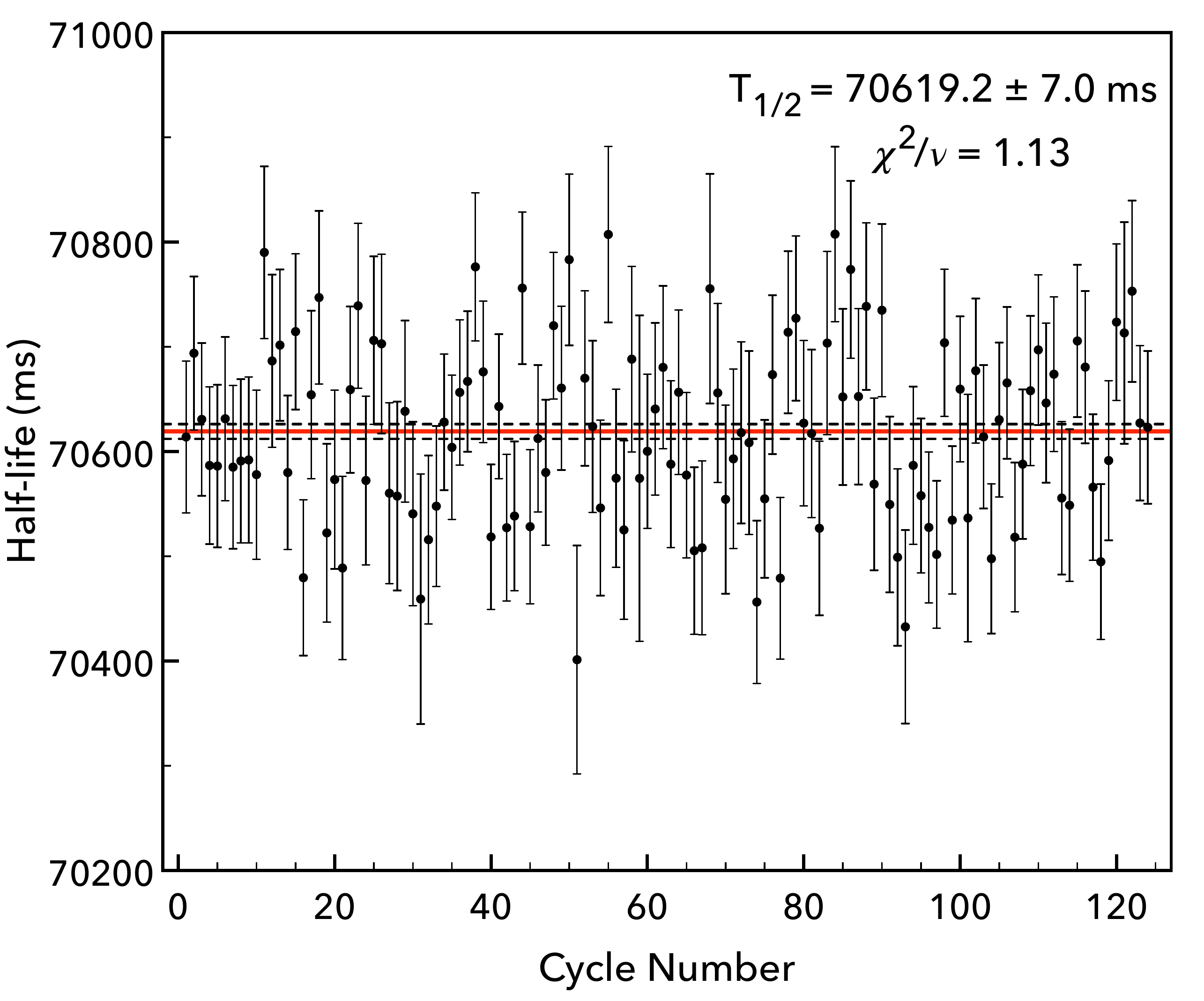}}
\caption{(Color online) Deduced half-life versus cycle number for the 124 decay cycles collected and analyzed in the present experiment.  A weighted average of them yields $T_{1/2}$($^{14}$O)~=~70619.2(70)~ms (solid and dashed lines) with a $\chi^2$/$\nu$~value of 1.13 for 123 degrees of freedom. }
\label{Half-life_vs_cycle}
\end{figure}

The MCS averaged half-lives obtained for all 124 decay cycles are plotted in Fig.~\ref{Half-life_vs_cycle}.  A weighted average of these values yields $T_{1/2}$~=~70619.2(70)~ms with a $\chi^2$/$\nu$ of 1.13 for 123 degrees of freedom.  This result is in excellent agreement with the previous world average of this quantity $T_{1/2}$~=~70619(11)~ms~\cite{Har20}.

\subsection{Electronic and Experimental Settings}

In order to search for possible sources of systematic uncertainty in the final result, the electronic settings used during the $^{14}$O data collection were modified on a cycle-by-cycle basis.  The 124 cycles, described above, were obtained with different detector voltage settings, different discriminator settings and with different bin times applied to the MCS data.  The two fixed and non-extendible dead times were swapped repeatedly throughout the course of the experiment and the beam-on time (initial activity collected) was also modified to search for potential rate-dependencies in the experimental data.

Before starting a new cycle, the discriminator threshold applied to the gas counter signals was varied, in 15~mV increments from 70~mV to 115~mV.  These thresholds were within the same range of values that were used to determine the plateau region (see Fig.~\ref{Plateau}).  Of the 124 cycles collected, 33 were obtained at the lowest threshold value of 70~mV, 29 were at 85~mV, 31 were at 100~mV and 31 were at 115~mV.  The half-life of $^{14}$O deduced from the weighted average of the cycles collected at each discriminator setting is presented in Table~\ref{Systematic_Table}. Because all 124 decay cycles were regrouped into 4 distinct threshold settings, the weighted average of these values is identical to the average of the 124 individual cycles.  Treating each of the 4 discriminator threshold settings as an independent measurement of the $^{14}$O half-life, yields a $\chi^2$/$\nu$~value of 1.02 (3 degrees of freedom).

\begin{figure}[tp]
\centering
\rotatebox{0}{\includegraphics[width=220pt]{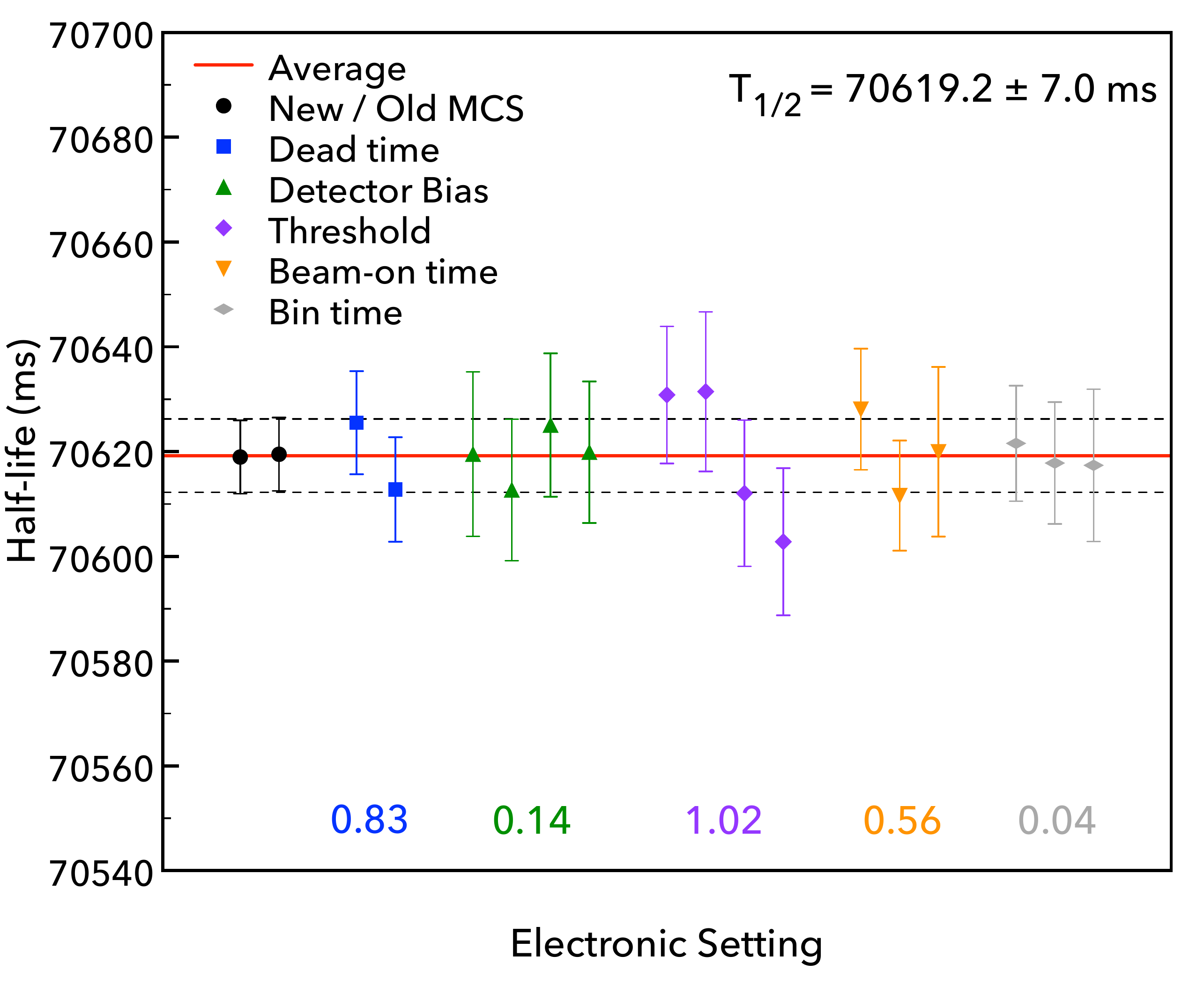}}
\caption{(Color online) Systematic plot that groups the half-lives deduced from the individual cycles data according to the particular electronic settings used in the experiment (see Table~\ref{Systematic_Table} for a detailed description of each point).  The solid and dashed lines correspond to the weighted average half-life obtained from the entire data set and its statistical uncertainty $T_{1/2}$~=~70619.2(70)~ms.  The $\chi^2$/$\nu$~value for each of the electronic settings is indicated. }
\label{Systematics}
\end{figure}

A similar procedure was performed for the other electronics settings in the experiment and the results are presented in Table~\ref{Systematic_Table} and in Fig.~\ref{Systematics}.  Four separate voltages were applied to the gas counter (3 degrees of freedom, $\chi^2$/$\nu$~=~0.14), four discriminator thresholds ($\chi^2$/$\nu$~=~1.02), three beam-on times ($\chi^2$/$\nu$~=~0.56), two non-extendible dead times ($\chi^2$/$\nu$~=~0.83) and three bin times ($\chi^2$/$\nu$~=~0.04) applied to the MCS.

All of the reduced $\chi^2$~values obtained from grouped sets of electronics settings are close to, or less than, unity. All of them are also smaller than the $\chi^2$/$\nu$ value of 1.13 obtained from the half-life deduced as a function of cycle number (Fig.~\ref{Half-life_vs_cycle}). While all of these values are consistent with fluctuations associated with statistical uncertainties, we nevertheless adopt the method of the Particle Data Group~\cite{Zyl20} to assign a conservative systematic uncertainty to the final result. With this method, the overall statistical uncertainty of 7.0~ms is multiplied by the square root of 1.13 (the largest reduced~$\chi^2$ value), which increases the total uncertainty to 7.5~ms.  If this total uncertainty is obtained from the quadrature sum of the statistical uncertainty with the systematic, then the half-life of $^{14}$O deduced in the present experiment is:
\begin{eqnarray}
T_{1/2}\mathrm{(}^{14}\mathrm{O)} = 70619.2(70)(26)~\mathrm{ms},
\label{half-life_withsyst}
\end{eqnarray}
where the first uncertainty is statistical and the second is the systematic uncertainty obtained from the method of the Particle Data Group applied to the deduced half-life as a function of cycle number (Fig.~\ref{Half-life_vs_cycle}).

\begin{table}[htp]
\caption{Average values of the $^{14}$O half-life deduced from grouping the experimental data according to the particular electronic setting used in the experiment.  The second column provides the number of cycles $N_{c}$ collected at each setting.  Because each group contains the same experimental data, the sum of all cycles for each group is 124 and the weighted average of each group is $T_{1/2}$($^{14}$O)~=~70619.2(70)~ms. }
\label{Systematic_Table}
\begin{center}
\begin{tabular}{lcc}
\hline
\hline
Electronic Setting  & ~~N$_{c}$~~ & ~~Half-life~[ms]~~ \\
\hline
New 3 $\mu$s / Old 4 $\mu$s  & 61 & 70625.5(100)  \\
New 4 $\mu$s / Old 3 $\mu$s  & 63 & 70612.8(100)  \\
\hline
Bias: 2300~V   & 22  & 70619.5(157) \\
Bias: 2350~V   & 34  & 70612.7(135) \\
Bias: 2400~V   & 33  & 70625.1(137) \\
Bias: 2450~V   & 35  & 70619.9(135) \\
\hline
Threshold: 70~mV    & 33  & 70630.8(131) \\
Threshold: 85~mV    & 29  & 70631.4(152) \\
Threshold: 100~mV   & 31  & 70612.0(140) \\
Threshold: 115~mV   & 31  & 70602.8(140) \\
\hline
Beam on: 0 to 60~s    & 44  & 70628.1(116) \\
Beam on: 60 to 120~s  & 56  & 70611.6(105)  \\
Beam on: 120 to 180~s & 24  & 70620.0(162) \\
\hline
Bin time: 2.8~s & 48  & 70621.6(110) \\
Bin time: 3.0~s & 43  & 70617.8(116) \\
Bin time: 3.2~s & 33  & 70617.4(145) \\
\hline
\hline
\end{tabular}
\end{center}
\end{table}

\subsection{\label{isobaric_systematic}Search for Isobaric Contamination}

As described above, the $^{14}$O activity was delivered to the experiment as a $^{12}$C$^{14}$O$^+$ beam at the mass setting $A/q$~=~26. While the resolution of the mass separator at ISAC is sufficient to reject all elements and isotopes produced with $A/q$~$\neq$~26, it cannot completely eliminate all of the neighbouring $A/q$~=~26 isobars.  From the fit presented in Fig.~\ref{Run309}, it is already clear that significant amounts of contamination can be ruled out because the experimental data are well described by a single exponential decay.  However, because even trace amounts of isobaric contaminants have the potential to affect a high-precision half-life determination, it is of crucial importance to evaluate what effect their possible presence may have on the final result.

The complete list of isobars at $A$~=~26 consists of $^{26}$P, $^{26}$Si, $^{26}$Al, $^{26}$Mg, $^{26}$Na, $^{26}$Ne and $^{26}$F.  As $^{26}$Mg is stable, this isotope cannot produce any counts in the detector and so its presence cannot affect the measurement.  Ignoring the production yields, and even the mass differences for the remaining unstable $A$~=~26 isobars, several of these can also be considered completely negligible based solely on their half-lives.  As described above, every sample collection (beam on) period was followed by a 22.2~s waiting period before moving the sample into the gas detector.  Any radioactive species with a half-life of 1~s or less, would thus experience a waiting time that is more than 20~times its own half-life.  For example, if there was a beam contaminant with a half-life of 1~s, the fraction of activity remaining after this 22.2~s period compared to the initial activity implanted is only 2$\times$10$^{-7}$.  This eliminates $^{26}$P, $^{26}$Na, $^{26}$Ne and $^{26}$F as potential contaminants as they all have half-lives that are $<$1.1~s.  Furthermore, based on previous measurements at ISAC, yields of silicon isotopes (like $^{26}$Si) are heavily suppressed when using a silicon-carbide target since the diffusion and effusion probabilities of silicon through a silicon target are extremely low.  The yield of $^{10}$C$^{16}$O$^+$ from silicon carbide is also expected to be negligible since this would require significant amounts of oxygen to be present in the target.  This beam has been produced at ISAC in the past, but only from an oxide target~\cite{Dun16}.

This leaves $^{26}$Al as the only possible isobaric contaminant in the beam that may still be present in the data after the 22.2~s ``cool-down" time. The 5$^+$ ground-state of $^{26}$Al is radioactive, but it is very long lived with a half-life of $T_{1/2}$($^{26}$Al)~=~7.16$\times$10$^5$~years~\cite{Sam72}.  Because this is so much longer than the 25~min measurement time of each cycle, any $^{26}$Al activity would appear constant in the decay spectra and would therefore already be included in the free background parameter of Eqn.~\ref{fit_function}.  However, there is a 0$^+$ isomeric state in $^{26}$Al located at 228~keV above the 5$+$~ground state that also decays via a superallowed Fermi $\beta$~transition with a well-known half-life of $T_{1/2}$~=~6.34602(54)~s~\cite{Har20}. 

To determine by how much the $^{14}$O half-life deduced in the present experiment may be affected by trace amounts of $^{26\mathrm{m}}$Al isobaric contamination in the beam, the experimental data were re-fit using a fit function that included $^{14}$O decay, the overall constant background as well as an additional exponential decay component:
\begin{eqnarray}
y_{fit}(t) = a_1e^{-a_2t} + a_4e^{-a_5t} + a_3.
\label{fit_function_2}
\end{eqnarray}
In this expression, $a_1$ to $a_3$ are identical to Eqn.\ref{fit_function}, $a_4$ is the initial activity of the potential $^{26\mathrm{m}}$Al contaminant and $a_5$ is its decay constant.  The half-life of $^{26\mathrm{m}}$Al has previously been determined to high-precision and thus $a_5$ was fixed in the analysis.  All 124 decay cycles were re-fit to the function defined in Eqn.~\ref{fit_function_2} with the $^{26\mathrm{m}}$Al initial activity $a_4$ treated as a free parameter.

The weighted average of the 124 initial $^{26\mathrm{m}}$Al activities that were deduced in this analysis was $-$2.9$\pm$4.5~c/s, which is consistent with zero.  Based on the known mass difference between $^{26\mathrm{m}}$Al and $^{12}$C$^{14}$O and the resolution of the mass separator at ISAC together with the 22.2~s ``cool-down" time, this was expected to be very small and this analysis provides further confirmation.  The conclusion is that $^{26\mathrm{m}}$Al isobaric contamination present in the $^{14}$O decay data can be considered negligible.    

\subsection{\label{Chop}Leading Channel Removal Analysis}

\begin{figure}[tp]
\centering
\rotatebox{0}{\includegraphics[width=220pt]{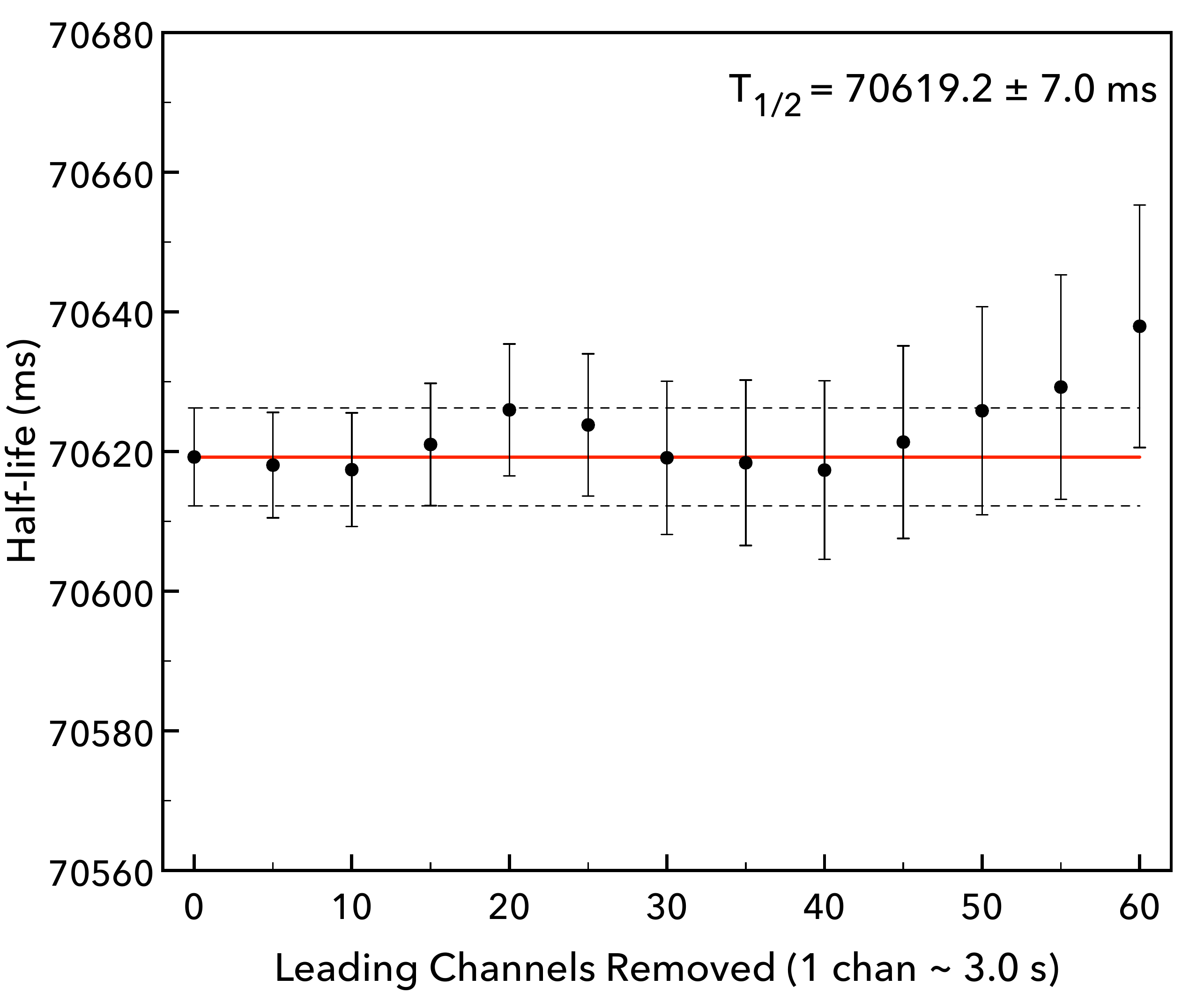}}
\caption{(Color online) Half-life of $^{14}$O versus the leading number of channels removed from the analysis.  The weighted average $T_{1/2}$~=~70619.2(70)~ms of the 124 cycles with no channels removed (solid and dashed lines) is overlayed for comparison.  As these data are not statistically independent, a weighted average of them cannot be determined. }
\label{Chopplot}
\end{figure}

As a final test to search for any other rate-dependent effects in the data, the half-life of $^{14}$O was determined as a function of the starting time of the fit to the exponential decay activity curves.  In this analysis, data from the leading channels were systematically removed from the fit in steps of 5 channels (average of 15~s/channel) and the $^{14}$O half-life was determined, as described above, from the weighted average of the 124 cycles.  The result of this analysis is presented in Fig.~\ref{Chopplot}.  The first point, which corresponds to no leading channels removed, is the weighted average $T_{1/2}$~=~70619.2(70)~ms that was determined above.  Removing up to 60~channels from the start of the decay curves (approximately 180~s or 2.5~times the half-life of $^{14}$O) shows that the deduced half-life is independent of the initial counting rate.  This confirms that any and all additional sources of rate-dependent effects that were not accounted for in the analysis, including the possibility that trace amounts of contaminants may be present, are negligible.

\subsection{\label{deadtime_systematic}Uncertainty in the Measured Dead Times}

As described above, the fixed and non-extendible dead times that were applied to the new and old MCS modules and were used to correct the experimental data were determined to be 2.9958(42)~$\mu$s and 4.0105(43)~$\mu$s using the source-plus-pulser method.  Their central values were used to correct the $^{14}$O decay data but the impact of their uncertainties on the $^{14}$O half-life still needs to be evaluated.  To accomplish this, the $^{14}$O decay data were corrected and re-fit using the dead times of 3.0000~$\mu$s and 4.0148~$\mu$s (nominal values +~1$\sigma$) and again using 2.9916~$\mu$s and 4.0062~$\mu$s (nominal $-$~1$\sigma$).  The half-life of $^{14}$O obtained from adjusting the fixed dead times within one standard deviation of their central values, resulted in a shift to its central value by a maximum of 1.4~ms.  This is negligible compared to the $\pm$~7.0~ms statistical uncertainty in the $^{14}$O half-life measurement, but it was nevertheless included as an additional source of uncertainty in the final result. 

\subsection{Final half-life result}

A summary of all of the uncertainties (statistical and systematic) that were used to evaluate the total uncertainty in the $^{14}$O half-life are summarized in Table~\ref{error_budget}.  The two largest contributions are the $\pm$~7.0~ms statistical uncertainty and the systematic uncertainty of $\pm$~2.6~ms, that was assigned to account for the $\chi^2$/$\nu$~value of 1.13 obtained from the half-life determined as a function of cycle number.  Summing all of these uncertainties in quadrature leads to an overall uncertainty of $\pm$~7.6~ms.  The half-life of $^{14}$O deduced in the present work is therefore: 
\begin{eqnarray}
T_{1/2}\mathrm{(}^{14}\mathrm{O)} = 70619.2 \pm 7.6~\mathrm{ms}.
\label{half-life_final}
\end{eqnarray}
This result has been determined at the level of $\pm$~0.011$\%$ and is the most precise half-life ever determined for $^{14}$O.  It is also in excellent agreement with the previous world-average, $T_{1/2}$~=~70619(11)~ms, obtained from 11 previous measurements of this quantity~\cite{Har20}. 

\begingroup
\begin{table}[tp]
\caption{Uncertainty budget in the analysis of the half-life of $^{14}$O deduced in the present experiment. }
\label{error_budget}
\begin{center}
\begin{tabular}{lc}
\hline
\hline
Source  &  Uncertainty (ms)   \\
\hline
Statistical uncertainty                             & 7.0  \\
Reduced $\chi^2$ (Fig.~\ref{Half-life_vs_cycle})    & 2.6  \\
Measured dead times                                 & 1.4  \\
\hline
Total (quadrature sum)                              & 7.6 \\
\hline
\hline
\end{tabular}
\end{center}
\end{table}
\endgroup

\begin{figure}[tp]
\centering
\rotatebox{0}{\includegraphics[width=220pt]{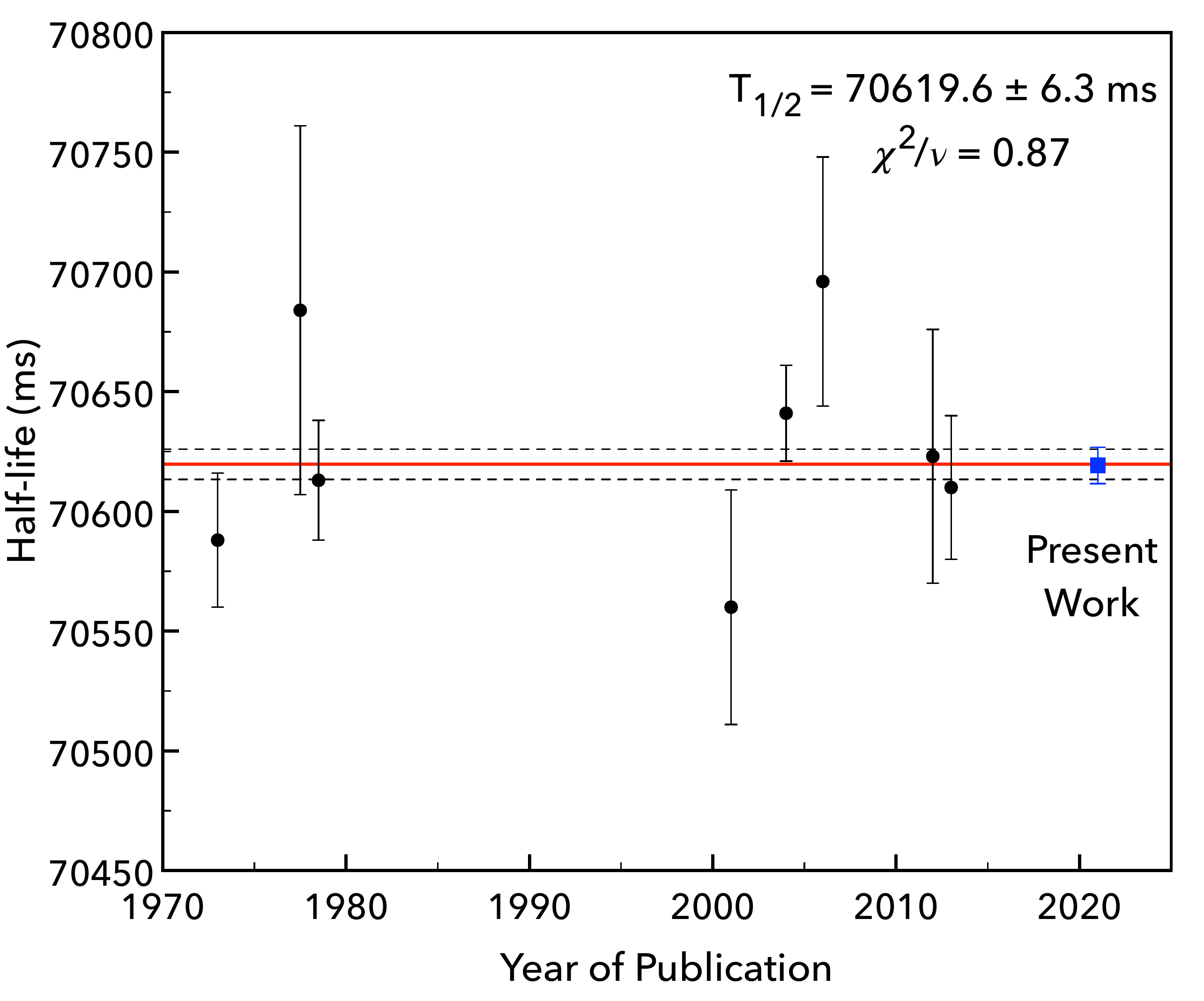}}
\caption{(Color online) Comparison of the present result with the previous most precise measurements of the half-life of $^{14}$O from Refs.~\cite{Cla73,Bec78,Wil78,Gae01,Bar04,Bur06,Tak12,Laf13}. The red solid line is the new world average that includes the present result and the dashed lines represent the $\pm$1$\sigma$ uncertainties.  The $\chi^2$/$\nu$~value for the set is 0.87 for 8 degrees of freedom.}
\label{halflife_worldaverage}
\end{figure}

The half-life of $^{14}$O deduced in the present work is compared, in Fig.~\ref{halflife_worldaverage}, with the previous most precise measurements of this quantity~\cite{Cla73,Bec78,Wil78,Gae01,Bar04,Bur06,Tak12,Laf13}.  The procedure described in Ref.~\cite{Har20} for calculating a new world average $^{14}$O half-life was adopted. This method retains only those measurements whose quoted uncertainties are no more than 10 times larger than the most precise measurement.  The weighted average of the current result with the eight previous measurements that satisfy this criterion is: 
\begin{eqnarray}
T_{1/2}\mathrm{(}^{14}\mathrm{O)} = 70619.6 \pm 6.3~\mathrm{ms}.
\label{half-life_world}
\end{eqnarray}
The $\chi^2$/$\nu$~value for the 9 measurements plotted in Fig.~\ref{halflife_worldaverage} is 0.87, which indicates a consistent set.  The $^{14}$O half-life has now been established at the level of $\pm$~0.0089$\%$, which is now one of the most precise half-lives ever determined for any radioactive decay. 

In Ref.~\cite{Laf13}, a possible systematic discrepancy in the $^{14}$O world half-life data associated with the type of activity measurement (direct $\beta$~counting or $\gamma$-ray photopeak counting) was described that was based on the $\chi^2$/$\nu$ value of 2.83 obtained when the previous results were grouped according to the type of decay radiation detected.  Adding the direct $\beta$~counting result determined in the present work into this analysis yields the average values of $T_{1/2}$($\beta$)~=~70621(7)~ms (from 4 total measurements) and $T_{1/2}$($\gamma$)~=~70603(16)~ms (5~measurements) for direct $\beta$~and $\gamma$-ray photopeak counting, respectively.  The $\chi^2$/$\nu$~value for these two values is 1.26, which is consistent with there being no significant source of systematic uncertainty in the world data that can be attributed to the type of radiation detected.

\subsection{The $ft$ and $\mathcal{F}t$ values for $^{14}$O decay}

In terms of the $ft$~value for the superallowed decay of $^{14}$O, the new world average $^{14}$O half-life leads to an updated value of $ft$~=~3042.25(73)~s. This is a slight improvement compared to the value of $ft$~=~3042.23(84)~s from the most recent survey of superallowed data~\cite{Har20}.  Similarly, the corrected $\mathcal{F}t$~value for $^{14}$O is updated to the new value of $\mathcal{F}t$~=~3070.3(19)~s, which is almost unchanged from its previous value $\mathcal{F}t$~=~3070.2(19)~s~\cite{Har20}.  The reason for these small differences is because the new world average half-life is numerically almost identical to the previous average and because uncertainties in the $Q$~value and the branching ratio are much larger contributors to the overall uncertainty in the $ft$~value. As a result, the high-precision half-life that was deduced in the present measurement is not expected to impact the limit obtained in Ref.~\cite{Har20} on the possible presence of scalar currents.

\section{Conclusion}

The half-life of $^{14}$O that was deduced in the present work, $T_{1/2}$~=~70619.2(76)~ms, is the most precise value ever obtained for this nucleus and is more precise than the previous world average that was deduced from a weighted average of 11 independent measurements.  The new world average is entirely consistent with the previous one but the overall uncertainty in the half-life has been reduced from $\pm$~11.2 ms to $\pm$~6.3 ms, or nearly a factor of~2. Future improvements in the $ft$~value and the corrected $\mathcal{F}t$~value for the superallowed decay of $^{14}$O to improve upon existing limits on scalar interactions in nuclear $\beta$~decay can only be achieved through improved measurements of the $Q$~value and branching ratio for this decay and would require a reduction in the theoretical uncertainty associated with isospin-symmetry breaking. 

\section*{Acknowledgments}
We would like to thank the ISAC ion-source and beam development group for their hard work and dedication to the production and delivery of the high-quality $^{12}$C$^{14}$O beam that was necessary for this experiment.  We acknowledge the support of the Natural Sciences and Engineering Research Council of Canada (NSERC). Nous remercions le Conseil de recherches en sciences naturelles et en génie du Canada (CRSNG) de son soutien.  This work was supported by the U.S. Department of Energy, Office of Science under grant nos.~DE-FG02-93ER40789 and DE-SC-0017649.  CES acknowledges support from the Canada Research Chairs Program.  TRIUMF receives federal funding via a contribution agreement through the National Research Council of Canada.  The University of Regina is situated on the territories of the n\^{e}hiyawak, Anih\v{s}in\={a}p\={e}k, Dakota, Lakota, and Nakoda, and the homeland of the M\'{e}tis/Michif Nation and TRIUMF is situated in the traditional, ancestral and unceded territory of the Musqueam people and we acknowledge that our research was conducted on their lands.

\bibliographystyle{ieeetr}
\bibliography{14O_EPJ.bib}

\begin{thebibliography}{10}

\bibitem{Har20}
{J.C.~Hardy and I.S.~Towner} {\em Phys. Rev. C}, vol.~102, p.~045501, 2020.

\bibitem{Fey58}
{R.P.~Feynman and M.~Gell-Mann} {\em Phys. Rev.}, vol.~109, p.~193, 1958.

\bibitem{Dun16}
{M.R.~Dunlop $et$~$al.$} {\em Phys. Rev. Lett.}, vol.~116, p.~172501, 2016.

\bibitem{Val15}
{A.A.~Valverde $et$~$al.$} {\em Phys. Rev. Lett.}, vol.~114, p.~232502, 2015.

\bibitem{Har15}
{J.C.~Hardy and I.S.~Towner} {\em Phys. Rev. C}, vol.~91, p.~025501, 2015.

\bibitem{Ame14}
{F.~Ames $et$~$al.$} {\em Rev. Sci. Instrum.}, vol.~85, p.~02B912, 2014.

\bibitem{Bal01}
{G.C.~Ball $et$~$al.$} {\em Phys. Rev.Lett.}, vol.~86, p.~1454, 2001.

\bibitem{Gri05}
{G.F.~Grinyer $et$~$al.$} {\em Phys. Rev. C}, vol.~71, p.~044309, 2005.

\bibitem{Gri08}
{G.F.~Grinyer $et$~$al.$} {\em Phys. Rev. C}, vol.~77, p.~015501, 2008.

\bibitem{Fin11}
{P.~Finlay $et$~$al.$} {\em Phys. Rev. Lett.}, vol.~106, p.~032501, 2011.

\bibitem{Bae65}
A.~Baerg {\em Metrologia}, vol.~131, p.~1, 1965.

\bibitem{Kos97}
{V.T.~Koslowsky $et$~$al.$} {\em Nucl. Instrum. Meth. Phys. Res. A}, vol.~401,
  p.~289, 1997.

\bibitem{Zyl20}
{P.A.~Zyla $et$~$al.$} {\em Prog. Theor. Exp. Phys.}, vol.~2020, p.~083C01,
  2020.

\bibitem{Sam72}
{E.A.~Samworth, E.K.~Warburton and G.A.P.~Engelbertink} {\em Phys. Rev. C},
  vol.~5, p.~138, 1972.

\bibitem{Cla73}
{G.J.~Clark, J.M.~Freeman, D.C.~Robinson, J.S.~Ryder, W.E.~Burcham and
  G.T.A.~Squier} {\em Nucl. Phys. A}, vol.~215, p.~429, 1973.

\bibitem{Bec78}
{J.A.~Becker, R.A.~Chalmers, B.A.~Watson and D.H.~Wilkinson} {\em Nucl.
  Instrum. Meth. Phys. Res.}, vol.~155, p.~211, 1978.

\bibitem{Wil78}
{D.H.~Wilkinson, A.~Gallmann and D.E.~Alburger} {\em Phys. Rev. C}, vol.~18,
  p.~401, 1978.

\bibitem{Gae01}
{M.~Gaelens $et$~$al$} {\em Eur. Phys.J A}, vol.~11, p.~413, 2001.

\bibitem{Bar04}
{P.H.~Barker, I.C.~Barnett, G.J.~Baxter and A.P.Byrne} {\em Phys. Rev.C},
  vol.~70, p.~024302, 2004.

\bibitem{Bur06}
{J.T.~Burke, P.A.~Vetter, S.J.~Freedman, B.K.~Fujiwara and W.T.~Winter} {\em
  Phys. Rev. C}, vol.~74, p.~025501, 2006.

\bibitem{Tak12}
{V.T.~Takau, M.N.~Thompson, R.J.~Scott, R.P.~Rassool and G.J.~O'Keefe} {\em
  Rad. Phys. Chem.}, vol.~81, p.~1669, 2012.

\bibitem{Laf13}
{A.T.~Laffoley $et$~$al.$} {\em Phys. Rev. C}, vol.~88, p.~015501, 2013.

\end{thebibliography}

\end{document}